\documentclass[12pt]{article}

\usepackage{bbm}
\usepackage{epsfig}
\usepackage{array}
\usepackage{float}
\usepackage{dsfont}
\usepackage{amstext}
\usepackage{rotating}
\usepackage{a4}
\usepackage{a4wide}
\usepackage{cite}
\usepackage{amsmath} 

\def\ra{\rightarrow}
\def\be{\begin{equation}}
\def\ee{\end{equation}}
\def\gs{\mathrel{
   \rlap{\raise 0.511ex \hbox{$>$}}{\lower 0.511ex \hbox{$\sim$}}}}
\def\ls{\mathrel{
   \rlap{\raise 0.511ex \hbox{$<$}}{\lower 0.511ex \hbox{$\sim$}}}}

\newcommand{\ba}{\begin{array}{c}}
\newcommand{\baz}{\begin{array}{cc}}
\newcommand{\barrr}{\begin{array}{rrr}}
\newcommand{\bad}{\begin{array}{ccc}}
\newcommand{\bav}{\begin{array}{cccc}}
\newcommand{\baf}{\begin{array}{ccccc}}
\newcommand{\bea}{\begin{equation} \begin{array}{c}}
\newcommand{\eea}{ \end{array} \end{equation}}
\newcommand{\ea}{\end{array}}
\newcommand{\D}{\displaystyle}

\newcommand{\eps}{\epsilon}

\newcommand{\gsim}{\raise0.3ex\hbox{$\;>$\kern-0.75em\raise-1.1ex\hbox{
   $\sim\;$}}} 
\newcommand{\lsim}{\raise0.3ex\hbox{$\;<$\kern-0.75em\raise-1.1ex\hbox{
   $\sim\;$}}}

\begin{document}

\title{\vspace{-1cm}
\hfill {\small SISSA 32/2010/EP}\\
\vskip 1.8cm
\bf \Large
On Leptonic Unitary Triangles and Boomerangs}
\author{
~~Alexander Dueck$^a$\thanks{email:
\tt alexander.dueck@mpi-hd.mpg.de}~\mbox{
},~~
Serguey T.~Petcov$^b$\thanks{Also at: INRNE, 
Bulgarian Academy of Sciences, 1784 Sofia, Bulgaria.
}\mbox{ },~~
Werner Rodejohann$^a$\thanks{email: 
\tt werner.rodejohann@mpi-hd.mpg.de}
\\\\
{\normalsize \it$^a$Max--Planck--Institut f\"ur Kernphysik,}\\
{\normalsize \it  Postfach 103980, D--69029 Heidelberg, Germany}  \\
\\ 
{\normalsize \it $^b$SISSA and INFN - Sezione di Trieste,}\\
{\normalsize \it Via Bonomea 265, I--34136 Trieste, Italy,}\\
{\normalsize and}\\
{\normalsize \it 
IPMU, University of Tokyo, Tokyo, Japan} 
}

\date{}
\maketitle
\thispagestyle{empty}
\vspace{0.8cm}
\begin{abstract}
\noindent  
We review the idea of leptonic unitary triangles and extend 
the concept of the recently proposed unitary boomerangs 
to the lepton sector. Using a convenient parameterization 
of the lepton mixing, we provide 
approximate expressions for the side lengths and 
the angles of the six different triangles and 
give examples of leptonic unitary boomerangs. 
Possible applications of the leptonic unitary boomerangs 
are also briefly discussed. 
\end{abstract}

\newpage
\noindent 
The leptonic, or Pontecorvo-Maki-Nakagawa-Sakata (PMNS), mixing matrix
$U$ can be written as 
\be \label{eq:U}
U = \left( \bad 
c_{12}  \, c_{13} 
& s_{12} \, c_{13} 
& s_{13} \, e^{-i \delta}  \\ 
-s_{12} \, c_{23} 
- c_{12} \, s_{23} \, 
s_{13}  \, e^{i \delta} 
& c_{12} \, c_{23} - 
s_{12} \, s_{23} \, s_{13} 
\, e^{i \delta} 
& s_{23}  \, c_{13}  \\ 
s_{12}   \, s_{23} - c_{12} 
\, c_{23}  \, s_{13} \, e^{i \delta} & 
- c_{12} \, s_{23} 
- s_{12} \, c_{23} \, 
s_{13} \, e^{i \delta} 
& c_{23}  \, c_{13}  
\ea   
\right) P \,,
\ee
where $c_{ij} = \cos \theta_{ij},\ s_{ij} = 
\sin \theta_{ij}$ and $\delta$ is the unknown Dirac CP-violating 
phase. The two equally unknown Majorana phases 
\cite{BHP80,SchValle80} 
appear in $P = {\rm diag}(1,e^{i\phi_2},e^{i \phi_3})$. 
We will focus on the
implications of $\delta$ in this letter. Namely, we discuss some 
aspects of unitarity of $U$ and Dirac-like CP violation in the form of
unitarity triangles and boomerangs \cite{UB,UB1,UB2}. 

In the standard parametrization given above, 
$U$ is obtained by three consecutive rotations: 
\bea \label{eq:rot}
U = R_{23}(\theta_{23}) \, \tilde{R}_{13}(\theta_{13}; \delta) \, 
R_{12}(\theta_{12})\, ,  \mbox{where e.g., } \\[0.1in]
R_{12}(\theta_{12}) = 
\left( \bad 
c_{12} & s_{12} & 0 \\
-s_{12} & c_{12} & 0 \\
0 & 0 & 1 
\ea
\right) , \quad
\tilde{R}_{13}(\theta_{13}; \delta)  = 
\left( \bad 
c_{13} & 0 & s_{13} \, e^{-i \delta} \\
0 & 1 & 0 \\ 
-s_{13}  \, e^{i \delta} & 0 & c_{13} 
\ea
\right) .
\eea
One notes that at zeroth order lepton mixing is well described (see Table 
\ref{tab:angles}) by tri-bimaximal mixing (TBM) \cite{tbm}
\be
U_{\rm TBM} = \left( 
\bad 
\sqrt{\frac 23} & \sqrt{\frac 13} & 0 \\
-\sqrt{\frac 16} & \sqrt{\frac 13} & \sqrt{\frac 12} \\
\sqrt{\frac 16} & -\sqrt{\frac 13} & \sqrt{\frac 12}
\ea
\right) ,
\ee
or $U_{\rm TBM} = R_{23}(\pi/4) \, R_{12}(\theta_{\rm TBM})$, where 
$\sin^2 \theta_{\rm TBM} = \frac 13$. Accepting this rather economic
scheme as the zeroth order description, makes possible to parameterize the
PMNS matrix around $U_{\rm TBM}$, i.e.~\cite{PRW}
\be \label{eq:PRW}
U = R_{23}(\pi/4) \, U_\epsilon \, R_{12}(\theta_{\rm TBM}) \, ,\mbox{ where } 
U_\epsilon = R_{23}(\epsilon_{23}) \, 
\tilde{R}_{13}(\epsilon_{13}; \delta) 
\, R_{12}(\epsilon_{12})\,.
\ee
The commonly used neutrino mixing observables are then obtained as 
zeroth order terms given by their TBM-values, and corrections in terms 
of the small $\epsilon_{ij}$:  
\begin{eqnarray}
\sin^2 \theta_{12} & = & \frac 13 \left(\cos \epsilon_{12} + 
\sqrt{2} \, \sin  \epsilon_{12} \right)^2   \simeq 
\frac 13 \left(1 + 2\sqrt{2} \, \epsilon_{12} + \epsilon_{12}^2
\right)  , \\
\sin^2 \theta_{23} & = & \frac 12 \left(1 + \sin 2 \epsilon_{23} \right)
\simeq \frac 12 \left(1 + 2 \, \epsilon_{23} \right)  ,\\
U_{e3} & = & \sin \epsilon_{13} \, e^{-i \delta} \, , \\ 
J_{\rm CP} & = & \frac{1}{24} \, 
(2\sqrt{2} \, \cos 2\epsilon_{12} + \sin 2 \epsilon_{12} ) \, 
\cos 2 \epsilon_{23} \, \sin 2 \epsilon_{13} \, \cos \epsilon_{13} \,
\sin \delta \\ 
&  \simeq & \left(1  + \frac{\epsilon_{12}}{\sqrt{2}} 
 \right) \frac{\epsilon_{13}}{3\sqrt{2}}  \sin \delta \, .
\end{eqnarray}
We have expanded up to second order in the small parameters
$\epsilon_{ij}$ and in the last line 
have given the usual Jarlskog invariant,  
\be
J_{\rm CP} = {\rm Im}\left\{ U_{e1}^\ast \, U_{\mu 3}^\ast \, U_{e3} \,
U_{\mu 1} \right\} = 
\frac 18  \sin 2 \theta_{12} \, \sin 2 \theta_{23} \, \sin 2
\theta_{13} \, \cos \theta_{13} \, \sin \delta \, ,
\ee
which describes leptonic Dirac-like CP violation. 
The advantage of the 
parameterization in Eq.~(\ref{eq:PRW}) is that each $\theta_{ij}$ is corrected 
from its TBM-value by only one of the $\epsilon_{ij}$. 
\begin{table}[t]
\centering
\begin{tabular}{c|ccc}
\hline\hline \\[0.0in]
Parameter  & best-fit$^{+1 \sigma}_{-1 \sigma}$ & $2 \sigma$ & 
    $3 \sigma$ \\[0.1in] \hline \\[0.0in]
$\sin^2 \theta_{12}$ & 
    0.318$^{+0.019}_{-0.016}$ & 0.29-0.36 & 0.27-0.38 \\
$\sin^2 \theta_{23}$ & 
    0.500$^{+0.070}_{-0.060}$ & 0.39-0.63 & 0.36-0.67 \\
$\sin^2 \theta_{13}$ & 
    0.013$^{+0.013}_{-0.009}$ & $\le 0.039$ & $\le 0.053$ \\[0.1in] 
\hline \hline 
\end{tabular}
\caption{Mixing angles and their best-fit values, 
1$\sigma$, 2$\sigma$ and 3$\sigma$ ranges 
    \cite{Schwetz2010}.}
\label{tab:angles}
\end{table}

The angles in Table \ref{tab:angles} 
are related to the elements of the PMNS 
matrix in the following way:
\bea
\sin^2\theta_{13} = |U_{e3}|^2\,,~
\sin^2\theta_{23} = \frac{|U_{\mu 3}|^2}{1 - |U_{e3}|^2}\,,~
\cos^2\theta_{23} = \frac{|U_{\tau 3}|^2}{1 - |U_{e3}|^2}\,, \\
\cos^2\theta_{12} = \frac{|U_{e1}|^2}{1 - |U_{e3}|^2}\,,~~
\sin^2\theta_{12} = \frac{|U_{e2}|^2}{1 - |U_{e3}|^2}\,.
\label{th12}
\eea
%

Information about the elements 
$|U_{e1}|^2$, $|U_{e2}|^2$  and $|U_{e3}|^2$ 
is obtained in the experiments with solar neutrinos 
and reactor antineutrinos. The experiments with 
atmospheric neutrinos 
provide data on $|U_{\mu 3}|^2$,
$|U_{\tau 3}|^2$ and $|U_{e3}|^2$. 
These elements can be measured also 
in long baseline experiments with accelerator 
$\nu_{\mu}$ and $\bar{\nu}_{\mu}$ (MINOS, T2K, NO$\nu$A, etc.).
Some of these experiments will use very intense 
neutrino beams and will search for CP violating effects 
in neutrino oscillations. The magnitude of these effects is 
determined by \cite{PKSP3nu88} the rephasing invariant 
$J_{\rm CP}$, associated with the Dirac phase $\delta$.

  The PMNS matrix is unitary, $U U^\dagger = U^\dagger U = \mathbbm{1}$,
and the six off-diagonal entries of $U U^\dagger = \mathbbm{1}$ and 
$U^\dagger U = \mathbbm{1}$ define six unitary triangles 
(properties of leptonic unitary triangles have been studied e.g.~in
Refs.~\cite{FS,branco,other}). 
Three stem from the conditions $U_{\alpha i}^\ast  U_{\beta i} = 0$ (the 
column-, or ``$\underline{\alpha \beta}$-triangles''), and the
other three from $U_{\alpha i}^\ast  U_{\alpha j} = 0$ 
(the row-, or ``$\underline{ij}$-triangles''). To be 
more concrete: 
\begin{eqnarray}
\underline{e\mu} : & U_{e1}^\ast  U_{\mu 1} + U_{e2}^\ast  U_{\mu
2} + U_{e3}^\ast U_{\mu 3} = 0 \, , \label{eq:emu} \\ 
\underline{e\tau} : & U_{e1}^\ast U_{\tau 1} + U_{e2}^\ast  U_{\tau
2} + U_{e3}^\ast  U_{\tau 3} = 0 \, ,  \label{eq:etau}\\ 
\underline{\mu\tau} : & U_{\mu 1}^\ast U_{\tau 1} + U_{\mu 2}^\ast  U_{\tau
2} + U_{\mu 3}^\ast  U_{\tau 3} = 0 \, ,  \label{eq:mutau}\\
\underline{12} : & U_{e 1}^\ast U_{e 2} + U_{\mu 1}^\ast U_{\mu
2}  + U_{\tau 1}^\ast  U_{\tau 2} = 0 \, ,  \label{eq:12} \\
\underline{13} : & U_{e 1}^\ast  U_{e 3} + U_{\mu 1}^\ast  U_{\mu
3}  + U_{\tau 1}^\ast  U_{\tau 3} = 0 \, ,\label{eq:13} \\
\underline{23} : & U_{e 2}^\ast  U_{e 3} + U_{\mu 2}^\ast  U_{\mu
3}  + U_{\tau 2}^\ast  U_{\tau 3} = 0 \, . \label{eq:23}
\end{eqnarray}
All six triangles have a common area of $A = \frac 12 J_{\rm CP}$. 
If we rephase the rows of the PMNS matrix via 
$U_{\alpha i} \ra U_{\alpha i} \, e^{i \phi_i}$, then the 
$\underline{\alpha \beta}$-triangles are invariant, whereas the 
$\underline{ij}$-triangles are modified:  
\be
U_{\alpha i}^\ast  U_{\alpha j} = 0 \ra 
U_{\alpha i}^\ast  U_{\alpha j} \, e^{i (\phi_j - \phi_i)} = 0\, .
\ee 
The $\underline{ij}$-triangles are rotated in the complex plane. 
This shows that the 
Majorana phases influence the rotation of the 
$\underline{ij}$-triangles and hence the $\underline{ij}$-triangles 
are in principle probing them \cite{branco}: with
a suitable phase rotation one can always arrange one side of the
$\underline{\alpha \beta}$-triangles to lie on the $x$- or $y$-axis. If
neutrinos are Majorana particles and the Majorana CP phases do not
take on CP conserving values (i.e., $0$ mod $\pi/2$) then this is not
possible for the $\underline{ij}$-triangles \cite{branco}. 
In this letter we will focus on the Dirac phase only. 
\\ 

A triangle is defined by, e.g., two side lengths and one angle, 
which are three parameters, one short of the four parameters, 
$\theta_{12}$, $\theta_{13}$, $\theta_{23}$ and $\delta$, of the PMNS 
matrix, which enter into the expression for $J_{\rm CP}$ 
and are associated with the Dirac-like CP violation 
in the lepton sector.
However, as two triangles share one common angle, it is possible 
to form an object called ``boomerang'' from two triangles
\cite{UB,UB1,UB2}. The boomerang is described by four parameters. 
We will illustrate leptonic unitary boomerangs 
in the following, using the convenient parameterization 
from Eq.~(\ref{eq:PRW}) of the PMNS matrix around 
the tri-bimaximal mixing.  

The approximate form of the PMNS matrix, 
\be
U = {\cal O}\left( 
\bad 
1 & 1 & \lambda^n \\
1 & 1 & 1 \\
1 & 1 & 1 
\ea
\right), 
\ee
where $\lambda $ can be taken as the sine of the Cabibbo angle 
and the exponent $n \ge 1$ is currently unknown, implies that 
the $\underline{12}$- and $\underline{\mu \tau}$-triangles have sides
of similar magnitude regardless of the value of $|U_{e3}|$. To give a
more precise estimate of the side lengths, we use now $\theta_{23} =
\pi/4$, $\theta_{12} = \theta_{\rm TBM}$, and keep only non-zero
$\theta_{13} = \epsilon_{13}$. The side lengths of the six triangles 
are:  
\begin{eqnarray} \D 
\underline{e\mu} : & \frac 13 \, , \, \frac 13 \, ,  \,
\frac{\epsilon_{13}}{\sqrt{2}} \, , & \\ \D 
\underline{e\tau} : & \frac 13 \, , \, \frac 13 \, , \,
\frac{\epsilon_{13}}{\sqrt{2}} \, , & \\ \D 
\underline{\mu\tau} : & \frac 13 \, , \, \frac 16 \, , \,
\frac 12 \, , & \\  
 \underline{12} : & \frac{1}{3\sqrt{2} } \, , \, \frac{\sqrt{2}}{3} \,
, \, \frac{1}{3\sqrt{2} } \, , & \\ \D 
 \underline{13} : & \frac{1}{2\sqrt{3} } \, , \, \sqrt{\frac 23} \,
\epsilon_{13}  \, , \, \frac{1}{2\sqrt{3} } \, , & \\ \D 
\underline{23} : & \frac{1}{\sqrt{6} } \, , \, \frac{\epsilon_{13}}{\sqrt{3}}
\, , \, \frac{1}{\sqrt{6} } \, . &
\end{eqnarray}
Hence, the maximal difference between two sides amounts to a factor 
$\simeq 3$ ($\simeq 2$) for the $\underline{\mu
\tau}$-($\underline{12}$-)triangle. 
\begin{figure}[th]
\begin{center}
\epsfig{file=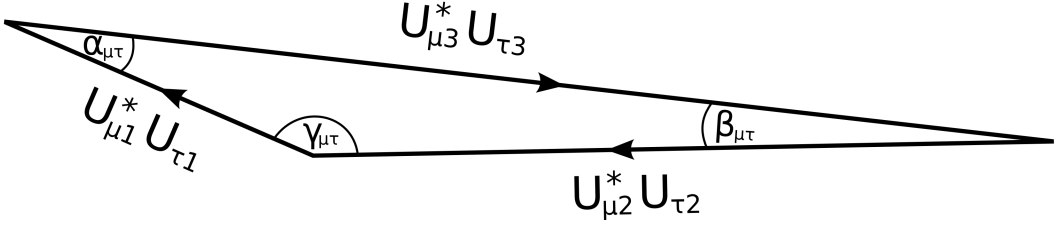,width=14cm}
\caption{\label{fig:ut_mt}
The $\underline{\mu\tau}$-triangle with best-fit values for the mixing
parameters (see Tab.~\ref{tab:angles}) 
and $\delta=\pi/2$. The angles are $\alpha_{\mu\tau} 
\simeq 19.0^\circ$, $\beta_{\mu\tau} \simeq 8.9^\circ$, 
and $\gamma_{\mu\tau} \simeq 152.1^\circ$, the side lengths $|U^\ast_{\mu2}U_{\tau2}| \simeq 0.34$, $|U^\ast_{\mu1}U_{\tau1}| \simeq 0.16$, and $|U^\ast_{\mu3}U_{\tau3}| \simeq 0.49$. The area is $A_{\underline{\mu\tau}} = 1/2 \, |U^\ast_{\mu2}U_{\tau2}| \, |U^\ast_{\mu3}U_{\tau3}| \sin \beta_{\mu\tau} \simeq 0.013$.}
\end{center}
\end{figure}
For the maximal allowed value of $\epsilon_{13}^{\rm max} = 0.23$ the
maximal difference for the other four triangles (from top to bottom)
is about 2, 2, 1.5 and 3. Taking the best-fit value of 
$\epsilon_{13}^{\rm best-fit} = 0.11$ yields values of about 
4, 4, 3 and 6. 
We will focus here for the sake of illustration 
on the $\underline{12}$- and $\underline{\mu \tau}$-triangles and
their resulting boomerang, and delegate the formulae for the remaining
four triangles to the Appendix.

Consider first the $\underline{\mu \tau}$-triangle 
(see Fig.~\ref{fig:ut_mt}): the lengths of the sides are 
\begin{eqnarray} \label{eq:mt1s}
|U_{\mu 2}^\ast U_{\tau 2}| & \simeq & \frac 13 \, \Bigl(
1 - \sqrt{2} \, \epsilon_{12} - \frac 12 \, \epsilon_{12}^2 - 2 \,
\epsilon_{23}^2 - 2 \sqrt{2} \, \epsilon_{13} \, \epsilon_{23} \, \cos
\delta - \frac 12 \, \epsilon_{13}^2 \, \cos 2 \delta \Bigr) , 
\\ \label{eq:mt2s}%
|U_{\mu 1}^\ast U_{\tau 1}| & \simeq & \frac 16 \left(1 + 2 \sqrt{2} \, \epsilon_{12} + \epsilon_{12}^2 - 2 \, \epsilon_{23}^2 + 4 \sqrt{2} \, \epsilon_{13} \, \epsilon_{23} \cos \delta - 2 \, \epsilon_{13}^2 \, \cos 2 \delta \right) \, , \\%
\label{eq:mt3s}
|U_{\mu 3}^\ast U_{\tau 3}| & \simeq & \frac 12  \left( 1 - \epsilon_{13}^2 - \, 2 \, \epsilon_{23}^2 \right) . 
\end{eqnarray}
%
%
The angles are 
\begin{align}
\label{eq:mt1a}
\alpha_{\mu \tau} & = \arg \left\{ 
- \frac{U_{\mu 3}^\ast U_{\tau 3}}{U_{\mu 1}^\ast U_{\tau 1}}
\right\} \simeq \left( 2 \sqrt{2} \, \epsilon_{13} \, - \, 6 \, \epsilon_{12} \, \epsilon_{13} \right) \sin \delta \, , \\ \label{eq:mt2a}%
\beta_{\mu \tau} & = \arg \left\{ 
- \frac{U_{\mu 2}^\ast U_{\tau 2}}{U_{\mu 3}^\ast U_{\tau 3}}
\right\} \simeq \left( \sqrt{2} \, \epsilon_{13} \, + \, 3 \, \epsilon_{12} \, \epsilon_{13} \right) \sin \delta \, , 
\\ \label{eq:mt3a}%
\gamma_{\mu \tau} & = \arg \left\{ 
- \frac{U_{\mu 1}^\ast U_{\tau 1}}{U_{\mu 2}^\ast U_{\tau 2}}
\right\} \simeq \, \left( -3 \sqrt{2} \, \epsilon_{13} \, + \, 3 \, \epsilon_{12} \, \epsilon_{13} \right) \sin \delta  
\, . 
\end{align}

The definitions of the angles are such that 
the sides are ordered as written 
in Eqs.~(\ref{eq:emu}--\ref{eq:23}) and $\alpha$ is the angle 
between the third and the first sides, $\beta$ between the second and 
third, and $\gamma$ between the first and second. 
With these definitions all the angles can be positive or negative 
and their sum can be equal either to $\pi$ or to ($-\pi$):
$\alpha_{\mu \tau} + \beta_{\mu \tau} + \gamma_{\mu \tau} 
= \pm \pi$\footnote[5]{This can be shown for instance by multiplying in 
$\alpha_{\mu \tau} = \arg \left\{ 
- \frac{U_{\mu 3}^\ast U_{\tau 3}}{U_{\mu 1}^\ast U_{\tau 1}}
\right\}$ the denominator and numerator with $U_{\mu 2}^\ast U_{\tau 2}$, 
performing simple manipulations and noting that $\arg\{-1 \} = \pm \pi$.}.  
If, for instance, we use the exact expressions in 
Eqs.~(\ref{eq:mt1a}--\ref{eq:mt3a}), 
the best-fit values 
given in Table \ref{tab:angles} and $\delta = \pi/2$, we get 
$\alpha_{\mu \tau} \simeq 19.0^\circ$, 
$\beta_{\mu \tau}\simeq 8.9^\circ$, 
$\gamma_{\mu \tau} \simeq 152.1^\circ$, and
$\alpha_{\mu \tau} + \beta_{\mu \tau} + \gamma_{\mu \tau} = \pi$.
For  $\delta = -\pi/2$, however, we find
$\alpha_{\mu \tau} \simeq -19.0^\circ$, 
$\beta_{\mu \tau}\simeq -8.9^\circ$, 
$\gamma_{\mu \tau} \simeq -152.1^\circ$, and
$\alpha_{\mu \tau} + \beta_{\mu \tau} + \gamma_{\mu \tau} = - \pi$. 
The approximate expressions for the angles that we give in this letter 
add up to zero. Depending on the signs and values 
of the $\epsilon_{ij}$ and of 
$\delta$, one has to add $\pi$ or $-\pi$ to one of 
the expressions. For instance, for 
the best-fit values and $\delta = \pi/2$ ($-\pi/2$) 
one would have to add $\pi$ ($-\pi$) to 
$\gamma_{\mu\tau}$ in Eq.~(\ref{eq:mt3a}). For the angles of the 
$\underline{12}$-triangle given below in Eqs.~(\ref{eq:12s1}--\ref{eq:12s3})
one would have to add $\pi$ ($-\pi$) to $\beta_{12}$.  
Similar comments are valid for the angles of the other unitarity 
triangles we will consider in what follows. 
We show the unitary $\underline{\mu \tau}$-triangle in Fig.~\ref{fig:ut_mt}
in the case of $\delta = \pi/2$ and for
the best-fit values from Table \ref{tab:angles}.

As we see, in the case of the $\underline{\mu \tau}$-triangle, 
the Dirac CP-violating phase $\delta$ does not coincide with any of the 
angles of the triangle.
For some of the other triangles, however, 
the Dirac phase $\delta$ corresponds
to certain angles: we have, for instance, 
$\delta \simeq \alpha_{e\mu}$ (see the Appendix).

  The $\underline{12}$-triangle side lengths are given by
\begin{align} \label{eq:12s1}
|U_{\mu1}^\ast U_{\mu2}| & \simeq \frac 1{3\sqrt{2}}\, \Bigl( 1 +
\frac 1{\sqrt{2}}\, \eps_{12} - 2\, \eps_{23} + \frac 1{\sqrt{2}}\,
\eps_{13} \, \cos \delta - 2 \, \eps_{12}^2 - \,\eps_{13}^2 \, \bigl(1 - \frac 94 \, \sin^2 \delta \bigr) \notag \\
& \hspace{1cm} - \sqrt{2}\,\eps_{12}\,\eps_{23} - 4 \,\eps_{12}\,\eps_{13}
\cos \delta  \Bigr) \, , \\
\label{eq:12s2} %
|U_{e1}^\ast U_{e2}| & \simeq \frac {\sqrt{2}}{3}\, \Bigl( 1 + \frac
1{\sqrt{2}}\, \eps_{12} - 2 \, \eps_{12}^2 - \eps_{13}^2 \Bigr) \, , \\
\label{eq:12s3} %
|U_{\tau1}^\ast U_{\tau2}| & \simeq \frac 1{3\sqrt{2}}\, \Bigl( 1 +
\frac 1{\sqrt{2}}\, \eps_{12} + 2\,\eps_{23} - \frac 1{\sqrt{2}}\,
\eps_{13} \, \cos \delta - 2 \, \eps_{12}^2 - \eps_{13}^2 \, 
\bigl(1 - \frac 94 \, \sin^2 \delta \bigr) \notag \\
& \hspace{1cm} + \sqrt{2}\,\eps_{12} \, \eps_{23} + 
4 \, \eps_{12} \, \eps_{13} \cos \delta  \Bigr) \, ,
\end{align}
\begin{figure}[t]
\begin{center}
\epsfig{file=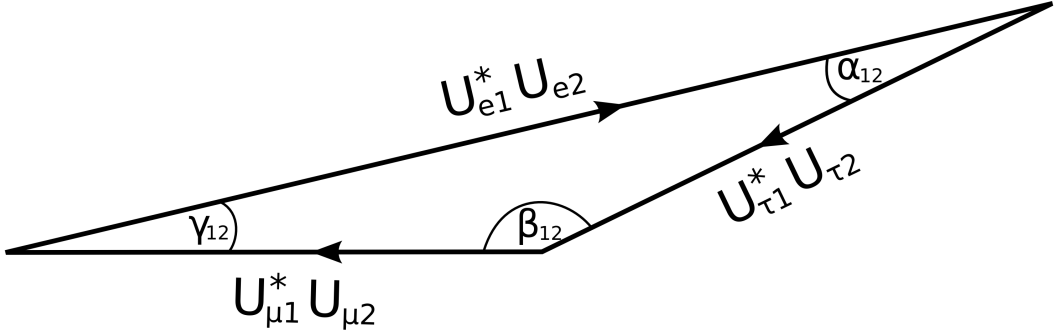,width=14cm}
\caption{\label{fig:ut_12}
The $\underline{12}$-triangle with best-fit values for the mixing
parameters and $\delta=\pi/2$. The angles are 
$\alpha_{12} \simeq 13.9^\circ$, 
$\beta_{12} \simeq 152.1^\circ$, and 
$\gamma_{12} \simeq 13.9^\circ$, the side lengths 
$|U^\ast_{\mu1}U_{\mu2}| \simeq 0.24$, $|U^\ast_{e1}U_{e2}| \simeq 0.46$, 
and $|U^\ast_{\tau1}U_{\tau2}| \simeq 0.24$. The area is 
$A_{\underline{12}} = 1/2 \, |U^\ast_{\mu1}U_{\mu2}| \, 
|U^\ast_{e1}U_{e2}| \sin \gamma_{12} \simeq 0.013$.
The $\underline{\mu\tau}$-triangle and the 
$\underline{12}$-triangle (Fig.~\ref{fig:ut_mt}) have one common angle: 
$\gamma_{\mu\tau} = \beta_{12}$.}
\end{center}
\end{figure}
and the angles by
\begin{align} \label{eq:12a1}
\alpha_{12} & = \arg \left\{ 
- \frac{U_{\tau 1}^\ast U_{\tau 2}}{U_{e 1}^\ast U_{e 2}}
\right\} \simeq \Bigl( \frac 3{\sqrt{2}} \, \eps_{13} \, - \, \frac 32 \,
\eps_{12} \, \eps_{13} \, - 
\, 3\sqrt{2} \, \eps_{13} \, \eps_{23} \Bigr) \sin \delta +
\frac 34\, \eps_{13}^2 \, \sin 2\delta \, , \\ \label{eq:12a2} %
\beta_{12} & = \arg \left\{ 
- \frac{U_{\mu 1}^\ast U_{\mu 2}}{U_{\tau 1}^\ast U_{\tau 2}} \right\} 
\simeq \Bigl( -3 \sqrt{2} \, \eps_{13} \, + \, 3 \, \eps_{12} \, \eps_{13}
\Bigr) \sin \delta  
\,, \\ \label{eq:12a3} %
\gamma_{12} & = \arg \left\{ 
- \frac{U_{e 1}^\ast U_{e 2}}{U_{\mu 1}^\ast U_{\mu 2}}
\right\} \simeq \Bigl( \frac 3{\sqrt{2}} \, \eps_{13} \, - \, \frac 32\, \eps_{12} \, \eps_{13} \, + \, 3\sqrt{2} \, \eps_{13}\, \eps_{23} \Bigr) \sin \delta - \frac 34 \, \eps_{13}^2 \, \sin 2\delta \, .
\end{align}
Using again the best-fit values 
from Table \ref{tab:angles} and $\delta = \pi/2$ we show 
the unitary $\underline{12}$-triangle in Fig.~\ref{fig:ut_12}. 

\begin{figure}[t]
\begin{center}
\epsfig{file=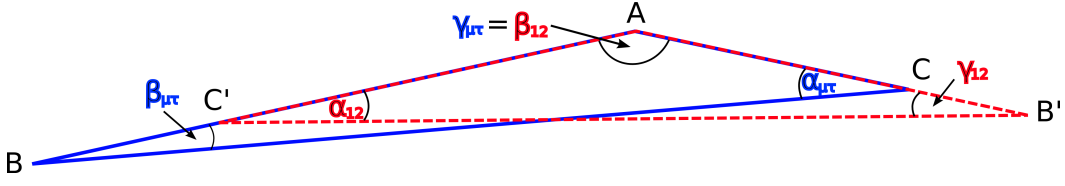,width=14cm}
\caption{\label{fig:ub_mt+12}
By overlapping the $\underline{\mu\tau}$-triangle from Fig.~\ref{fig:ut_mt} 
and the $\underline{12}$-triangle from Fig.~\ref{fig:ut_12} 
at their common angle we get the 
unitary $\underline{\mu\tau}$-$\underline{12}$-boomerang. 
}
\end{center}
\end{figure}

It is obvious from Eqs.~(\ref{eq:mt3a}) and (\ref{eq:12a2}) 
that $\gamma_{\mu\tau} = \beta_{12}$. This allows to 
``glue'' two triangles together and form an object called 
boomerang \cite{UB}. The $\underline{\mu\tau}$-$\underline{12}$-boomerang 
is shown in Fig.~\ref{fig:ub_mt+12}. 
The side lengths between the points are: 
AC  $= |U_{\mu1}^\ast U_{\tau1}| \simeq 0.16$,
AC' $= |U_{\tau1}^\ast U_{\tau2}| \simeq 0.24$,
AB  $= |U_{\mu2}^\ast U_{\tau2}| \simeq 0.34$,
AB' $= |U_{\mu1}^\ast U_{\mu2}| \simeq 0.24$,
BC  $= |U_{\mu3}^\ast U_{\tau3}| \simeq 0.49$ and
BC' $ = |U_{e1}^\ast U_{e2}| \simeq 0.46$.
Certain areas can be connected to the PMNS matrix elements and the
Jarlskog invariant \cite{UB2}. For instance, 
\be
A_{\rm ABB'} = \frac 12 \, \frac{|U_{\mu 2}|}{|U_{\tau 1}|} \, J_{\rm CP} \simeq 0.019~,~~
A_{\rm ACC'} = \frac 12 \, \frac{|U_{\tau 1}|}{|U_{\mu 2}|} \, J_{\rm CP} \simeq 0.009~,~~
\ee 
and thus $ A_{\rm ABB'} \, A_{\rm ACC'} 
= \frac 14 \, J_{\rm CP}^2
\simeq 1.7 \cdot 10^{-4}$. 
If there was some form of new physics which leaves $\gamma_{\mu\tau}$ 
and $\beta_{12}$ unaffected, 
then one might probe its presence by checking 
whether the product of the above areas is indeed 
given by $\frac 14 \, J_{\rm CP}^2$.\\

Unitary boomerangs are also a useful method to study the presence of 
new physics which one single unitary triangle might miss \cite{UB2}. 
Consider, for instance, the existence of a light sterile neutrino 
(generalization to more sterile species is straightforward). There 
would now be a unitary ``$\underline{\mu\tau}$-quadrangle'', defined as 
\[ 
 U_{\mu 1}^\ast U_{\tau 1} + U_{\mu 2}^\ast  U_{\tau
2} + U_{\mu 3}^\ast  U_{\tau 3} + U_{\mu 4}^\ast U_{\tau 4}= 0 \, .
\] 
In case of $U_{\tau 4}= 0$ (or $U_{\mu 4}= 0$), the experimental study of the 
$\underline{\mu\tau}$-triangle would reveal no anomaly. However, the 
$\underline{12}$-triangle might be modified, because with four generations 
one has now 
\[ 
U_{e 1}^\ast U_{e 2} + U_{\mu 1}^\ast U_{\mu
2}  + U_{\tau 1}^\ast  U_{\tau 2} = - U_{s 1}^\ast U_{s 2} \, . 
\]
Unless $U_{s 1} = 0$ or $U_{s 2} = 0$, an inconsistency between the 
two triangles would appear and 
the $\underline{\mu\tau}$-$\underline{12}$-boomerang could not be formed. 

   Unitary boomerangs are therefore a useful, illustrative and comprehensive 
method to study the consistence of the standard neutrino 
framework.

\vspace{0.3cm}
\begin{center}
{\bf Acknowledgments}
\end{center}
We would like to thank P.~Frampton for discussions
on unitary boomerangs. 
This work was supported by the ERC under the Starting Grant 
MANITOP and by the DFG in the Transregio 27 (A.D.~and W.R.), 
as well as by  World Premier International Research Center 
Initiative (WPI Initiative), MEXT, Japan, and
the INFN program on ``Astroparticle Physics'' 
(S.T.P.).

\renewcommand{\theequation}{A\arabic{equation}}
\setcounter{equation}{0}
\renewcommand{\thetable}{A\arabic{table}}
\setcounter{table}{0}

\begin{appendix}
\section{\label{sec:app}Formulae for the Remaining Unitary Triangles}
In the following we give the side lengths and angles 
for the remaining four unitary triangles. We will expand the formulae
in second order in the small $\epsilon_{ij}$. Recall that in the 
approximate expressions for the angles they add up to zero, and depending on 
the signs and values of the $\epsilon_{ij}$ and of $\delta$ one needs to 
add $\pm \pi$ for appropriate angles (see main text).

For the $\underline{e \mu}$-triangle one obtains the side lengths as 
\begin{align}
|U_{e2}^\ast U_{\mu 2}| & \simeq \frac 13 \, \Bigl( 1 + \frac
{\epsilon_{12}}{\sqrt{2}} - \epsilon_{23} - \frac 12  
\bigl( \sqrt{2} \, \epsilon_{13} + 4 \,  \epsilon_{12} \, \eps_{13} 
+ \sqrt{2} \, \eps_{13} \, \epsilon_{23} \bigr) \, \cos \delta \notag \\
& \hspace{1cm} - 2 \, \epsilon_{12}^2 - \frac {\epsilon_{12} \,
\epsilon_{23}}{\sqrt{2}} - \frac {\epsilon_{23}^2}{2} - 
\frac {\epsilon_{13}^2}{4} 
\, (2 - \sin^2 \delta)  \Bigr) \, , \\%
|U_{e 1}^\ast U_{\mu 1}| & \simeq \frac 13 \, \Bigl( 1 + \frac
{\epsilon_{12}}{\sqrt{2}} - \epsilon_{23} + \bigl( \sqrt{2} \,
\epsilon_{13} - 2 \, \epsilon_{12} \, \eps_{13} + \sqrt{2} \,
\eps_{13} \, \epsilon_{23} \bigr) \cos \delta \notag \\
& \hspace{1cm} - 2 \, \epsilon_{12}^2 - \frac {\epsilon_{12} \,
\epsilon_{23}}{\sqrt{2}} - \frac {\epsilon_{23}^2}{2} -
\frac 12 \, \epsilon_{13}^2 \, \cos 2 \delta \Bigr) \, , \\%
|U_{e3}^\ast U_{\mu 3}| & \simeq \frac { \eps_{13}}{\sqrt{2}} + \frac
{ \eps_{13} \, \eps_{23}}{\sqrt{2}}  \, ,
\end{align}
while the angles are 
\begin{align}
\alpha_{e\mu} & = \arg \left\{ 
- \frac{U_{e 3}^\ast U_{\mu 3}}{U_{e 1}^\ast U_{\mu 1}}
\right\} \simeq \delta - \left( \sqrt{2} \, \eps_{13} - \eps_{12} \, \eps_{13} \right) \sin \delta + \eps_{13}^2 \, \sin 2\delta \, , \\%
\beta_{e\mu} & = \arg \left\{ 
- \frac{U_{e 2}^\ast U_{\mu 2}}{U_{e 3}^\ast U_{\mu 3}}
\right\} \simeq - \, \delta - \Bigl( \frac {1}{\sqrt{2}} \, \eps_{13} - \frac 12 \, \eps_{12} \, \eps_{13} \Bigr) \sin \delta - \frac 14 \, \eps_{13}^2 \, \sin 2\delta \, , \\%
\gamma_{e\mu} & = \arg \left\{ 
- \frac{U_{e 1}^\ast U_{\mu 1}}{U_{e 2}^\ast U_{\mu 2}}
\right\} \simeq \Bigl(\frac 3{\sqrt{2}} \, \eps_{13} - \frac 32 \,
\eps_{12} \, \eps_{13} \Bigr) \sin \delta - \frac 34 \, \eps_{13}^2 \, \sin 2\delta  
\, .
\end{align}
Note that in the limit of $\epsilon_{13}$ going to zero (which 
corresponds to $|U_{e3}|$ going to zero and implies
that CP violation in neutrino oscillations is absent),
the exact expressions for $\alpha_{e\mu}$ and $\beta_{e\mu}$
are undefined, while the approximate expressions
for these two angles are formally given by $\delta$ and
$-\delta$, respectively. As $\delta$ is however unphysical when
$\epsilon_{13} = 0$, there is no inconsistency.
Turning to the $\underline{e \tau}$-triangle, we find 
\begin{align}
|U_{e2}^\ast U_{\tau 2}| & \simeq \frac 13 \, \Bigl( 1 + \frac 1{\sqrt{2}}\, \eps_{12} + \eps_{23} + \bigl( \frac 1{\sqrt{2}}\, \eps_{13} + 2 \eps_{12} \, \eps_{13} - \frac 1{\sqrt{2}}\, \eps_{13} \, \eps_{23} \bigr) \cos \delta \notag \\%
& \hspace{1cm} - 2\eps_{12}^2 - \frac 12\, \eps_{23}^2 + \frac 1{\sqrt{2}} \, \eps_{12} \, \eps_{23} - \frac 14 \, \eps_{13}^2 \, (2 - \sin^2 \delta) \Bigr) \, , \\%
|U_{e1}^\ast U_{\tau 1}| & \simeq \frac 13 \, \Bigl( 1 + \frac 1{\sqrt{2}}\, \eps_{12} + \eps_{23} + \bigl( -\sqrt{2} \, \eps_{13} + 2 \, \eps_{12} \, \eps_{13} + \sqrt{2} \, \eps_{13} \, \eps_{23} \bigr) \cos \delta \notag \\
& \hspace{1cm} - 2\eps_{12}^2 + \frac 1{\sqrt{2}} \, \eps_{12} \, \eps_{23} - \frac 12\, \eps_{23}^2 - \frac 12 \, \eps_{13}^2 \cos 2\delta \Bigr) \, , \\%
|U_{e3}^\ast U_{\tau 3}| & \simeq \frac 1{\sqrt{2}} \, \eps_{13} - \frac 1{\sqrt{2}} \, \eps_{13} \, \eps_{23} \, ,
\end{align}
and
\begin{align} 
\alpha_{e\tau} & = \arg \left\{ 
- \frac{U_{e 3}^\ast U_{\tau 3}}{U_{e 1}^\ast U_{\tau 1}} \right\}
 \simeq \delta + \Bigl( \sqrt{2} \, \eps_{13} - \eps_{12} \, \eps_{13} \Bigr) \sin \delta + \eps_{13}^2 \, \sin 2\delta \, , \\%
\beta_{e\tau} & = \arg \left\{ 
- \frac{U_{e 2}^\ast U_{\tau 2}}{U_{e 3}^\ast U_{\tau 3}}
\right\} \simeq - \, \delta + \Bigl( \frac{1}{\sqrt{2}} \, \eps_{13} - \frac 12 \, \eps_{12} \, \eps_{13} \Bigr) \sin \delta - \frac 14 \, \eps_{13}^2 \, \sin 2\delta \, , \\%
\gamma_{e\tau} & = \arg \left\{ 
- \frac{U_{e 1}^\ast U_{\tau 1}}{U_{e 2}^\ast U_{\tau 2}}
\right\} \simeq \Bigl( - \, \frac 3{\sqrt{2}} \, \eps_{13} + \frac 32 \, \eps_{12} \, \eps_{13} + 3\sqrt{2} \, \eps_{13} \, \eps_{23} \Bigr) \sin \delta - \frac 34\, \eps_{13}^2 \, \sin 2\delta  
\, .
\end{align}
For the $\underline{13}$-triangle we have
\begin{align}
|U_{\mu1}^\ast U_{\mu3}| & \simeq \frac 1{2\sqrt{3}}\, \Bigl( 1 + \sqrt 2 \, \eps_{12} + \sqrt 2 \, \eps_{13} \cos \delta - \frac 12 \, \eps_{12}^2 - 2 \, \eps_{23}^2 + \bigl( 2 \sqrt 2 \, \eps_{13} \, \eps_{23} - \eps_{12} \, \eps_{13} \bigr) \cos \delta \notag \\
& \hspace{1cm} - \frac 12 \eps_{13}^2 \cos 2\delta \Bigr) \, , \\%
|U_{e1}^\ast U_{e3}| & \simeq \sqrt{\frac{2}{3}}\, \eps_{13}\, - \frac{1}{\sqrt{3}}\, \eps_{12} \, \eps_{13} \, , \\%
|U_{\tau1}^\ast U_{\tau3}| & \simeq  \frac 1{2\sqrt{3}}\, \Bigl( 1 + \sqrt 2 \, \eps_{12} - \sqrt 2 \, \eps_{13} \cos \delta - \frac 12 \, \eps_{12}^2 - 2 \, \eps_{23}^2 + \bigl( \eps_{12} \, \eps_{13} + 2 \sqrt{2} \, \eps_{13} \,  \eps_{23} \bigr) \cos \delta \notag \\
& \hspace{1cm} - \frac 12 \eps_{13}^2 \cos 2\delta \Bigr) \, ,
\end{align}
and 
\begin{align}
\alpha_{13} & = \arg \left\{ 
- \frac{U_{\tau 1}^\ast U_{\tau 3}}{U_{e 1}^\ast U_{e 3}}
\right\} \simeq \delta + \Bigl( \sqrt{2} \, \eps_{13} - \eps_{12} \, \eps_{13} \Bigr) \sin \delta + \eps_{13}^2 \sin 2\delta \, , \\%
\beta_{13} & = \arg \left\{ 
- \frac{U_{\mu 1}^\ast U_{\mu 3}}{U_{\tau 1}^\ast U_{\tau 3}}
\right\} \simeq \Bigl( -2\sqrt{2} \, \eps_{13} \, + \, 6 \, \eps_{12} \, \eps_{13} \Bigr) \sin \delta  
\,, \\%
\gamma_{13} & = \arg \left\{ 
- \frac{U_{e 1}^\ast U_{e 3}}{U_{\mu 1}^\ast U_{\mu 3}}
\right\} \simeq -\delta + \Bigl( \sqrt 2 \, \eps_{13} - \eps_{12} \, \eps_{13} \Bigr) \sin \delta - \eps_{13}^2 \sin 2\delta \, .
\end{align}
Finally, we get for the $\underline{23}$-triangle
\begin{align}
|U_{\mu2}^\ast U_{\mu3}| & \simeq \frac 1{\sqrt{6}}\, \Bigl( 1 - \frac 1{\sqrt 2}\, \eps_{12} - \frac 1{\sqrt 2}\, \eps_{13} \cos \delta - \frac 12 \, \eps_{12}^2 - 2 \, \eps_{23}^2 \notag \\
& \hspace{1cm} - \bigl( \eps_{12} \, \eps_{13} + \sqrt 2 \, \eps_{13} \, \eps_{23} \bigr) \cos \delta - \frac 14\, \eps_{13}^2 \bigl(2 - \sin^2 \delta \bigr) \Bigr) \, , \\%
|U_{e2}^\ast U_{e3}| & \simeq \frac 1{\sqrt{3}}\, \eps_{13} + \sqrt{\frac 23}\, \eps_{12} \, \eps_{13} \, , \\%
|U_{\tau2}^\ast U_{\tau3}| & \simeq \frac 1{\sqrt{6}}\, \Bigl( 1 - \frac 1{\sqrt 2}\, \eps_{12} + \frac 1{\sqrt 2}\, \eps_{13} \cos \delta - \frac 12 \, \eps_{12}^2 - 2 \, \eps_{23}^2 \notag \\
& \hspace{1cm} + \bigl( \eps_{12} \, \eps_{13} - \sqrt 2 \, \eps_{13} \, \eps_{23} \bigr) \cos \delta - \frac 14\, \eps_{13}^2 \bigl(2 - \sin^2 \delta \bigr) \Bigr) \, ,
\end{align}
and for its angles
\begin{align}
\alpha_{23} & = \arg \left\{ 
- \frac{U_{\tau 2}^\ast U_{\tau 3}}{U_{e 2}^\ast U_{e 3}}
\right\} \simeq \delta - \Bigl( \frac 1{\sqrt{2}}\, \eps_{13} - \frac 12\, \eps_{12} \, \eps_{13} \Bigr) \sin \delta + \frac 14\, \eps_{13}^2 \sin 2\delta \, , \\%
\beta_{23} & = \arg \left\{ 
- \frac{U_{\mu 2}^\ast U_{\mu 3}}{U_{\tau 2}^\ast U_{\tau 3}}
\right\} \simeq \Bigl( \sqrt{2} \, \eps_{13} \, + \, 3 \, \eps_{12} \, \eps_{13} \Bigr) \sin \delta  
\, , \\%
\gamma_{23} & = \arg \left\{ 
- \frac{U_{e 2}^\ast U_{e 3}}{U_{\mu 2}^\ast U_{\mu 3}}
\right\} \simeq -\delta - \Bigl( \frac 1{\sqrt{2}}\, \eps_{13} - \frac 12 \, \eps_{12} \, \eps_{13} \Bigr) \sin \delta - \frac 14\, \eps_{13}^2 \sin 2\delta \, .
\end{align}
The following relations between the angles are found:
\begin{eqnarray}
\alpha_{e\mu} = -\gamma_{13} \, , 
\beta_{e \mu} = \gamma_{23} \, , 
\gamma_{e \mu} = \gamma_{12} \, , \\
\alpha_{e\tau} = \alpha_{13} \, ,  
\beta_{e \tau} = -\alpha_{23} \, , 
\gamma_{e \tau} = -\alpha_{12} \, , \\
\alpha_{\mu\tau} = -\beta_{13} \, ,  
\beta_{\mu \tau} = \beta_{23} \, , 
\gamma_{\mu \tau} = \beta_{12} \, .
\end{eqnarray}
There are only 9 different angles among the 18 in total. For each pair
of identical angles one can overlap the longer side of one triangle
with the shorter side of the other triangle or vice versa.   
There are therefore 18 possible unitary boomerangs \cite{UB}. 

\end{appendix}


\begin{thebibliography}{99}

\bibitem{BHP80}S.~M.~Bilenky, J.~Hosek and S.~T.~Petcov,
  Phys.\ Lett.\  B {\bf 94}, 495 (1980).



\bibitem{SchValle80} J.~Schechter and J.~W.~F.~Valle,
  Phys.\ Rev.\  D {\bf 22}, 2227 (1980); 
M.~Doi, T.~Kotani, H.~Nishiura, K.~Okuda and E.~Takasugi,
  Phys.\ Lett.\  B {\bf 102}, 323 (1981).





\bibitem{UB}P.~H.~Frampton and X.~G.~He,
 Phys.\ Lett.\  B {\bf 688}, 67 (2010)
  [arXiv:1003.0310 [hep-ph]].

\bibitem{UB1}
S.~W.~Li and B.~Q.~Ma,
  arXiv:1003.5854 [hep-ph].

\bibitem{UB2}P.~H.~Frampton and X.~G.~He,
  arXiv:1004.3679 [hep-ph].



\bibitem{tbm}
P.~F.~Harrison, D.~H.~Perkins and W.~G.~Scott,
  Phys.\ Lett.\  B {\bf 530}, 167 (2002)
  [arXiv:hep-ph/0202074];
Phys.\ Lett.\  B {\bf 535}, 163 (2002)
  [arXiv:hep-ph/0203209]; 
  Z.~Z.~Xing,
  Phys.\ Lett.\  B {\bf 533}, 85 (2002)
  [arXiv:hep-ph/0204049];
  X.~G.~He and A.~Zee,
  Phys.\ Lett.\  B {\bf 560}, 87 (2003)
  [arXiv:hep-ph/0301092].


\bibitem{PRW}S.~Pakvasa, W.~Rodejohann and T.~J.~Weiler,
  Phys.\ Rev.\ Lett.\  {\bf 100}, 111801 (2008)
  [arXiv:0711.0052 [hep-ph]].

\bibitem{PKSP3nu88}
P.~I.~Krastev and S.~T.~Petcov,
  Phys.\ Lett.\  B {\bf 205}, 84 (1988).



\bibitem{FS}
 Y.~Farzan and A.~Y.~Smirnov,
  Phys.\ Rev.\  D {\bf 65}, 113001 (2002)
  [arXiv:hep-ph/0201105].



\bibitem{branco}
 J.~A.~Aguilar-Saavedra and G.~C.~Branco,
  Phys.\ Rev.\  D {\bf 62}, 096009 (2000)
  [arXiv:hep-ph/0007025].

\bibitem{other} Z.~Z.~Xing and H.~Zhang,
  Phys.\ Lett.\  B {\bf 618}, 131 (2005)
  [arXiv:hep-ph/0503118]; 
J.~D.~Bjorken, P.~F.~Harrison and W.~G.~Scott,
  Phys.\ Rev.\  D {\bf 74}, 073012 (2006)
  [arXiv:hep-ph/0511201].



\bibitem{Schwetz2010}T.~Schwetz, M.~A.~Tortola and J.~W.~F.~Valle,
  New J.\ Phys.\  {\bf 10}, 113011 (2008)
  [arXiv:0808.2016v3 [hep-ph]].  

\end{thebibliography}
\end{document}